\documentclass{rspublic}
\usepackage{psfig}

\begin{document}

\title[X-ray Binaries]{Accretion Flows in 
X-ray Binaries}

\author[C. Done]{Chris Done}

\affiliation{Department of Physics, University of Durham, South Road,
Durham DH1 3LE, UK}

\label{firstpage}

\maketitle

\begin{abstract}{X-ray, Accretion, Low Mass X-ray Binaries, Black
Holes, Neutron stars}
I review the X-ray observations of Galactic accreting black holes and
neutron stars, and interpretions of these in terms of solutions of the
accretion flow equations. 
\end{abstract}

\section{Introduction}

One of the key puzzles in X-ray astronomy is to understand accretion
flows in a strong gravitational field. This applies to both Active
Galactic Nuclei (AGN) and Quasars, where the accretion is onto a
supermassive black hole, and to the stellar mass Galactic black holes
(GBH) and even the neutron star systems.  Neutron star radii are of
order three Schwarzchild radii, i.e. the same as that for the last
stable orbit of material around a black hole. Thus they have very
similar gravitational potentials so should have very similar accretion
flows, though of course with the major
difference that neutron stars then have a solid surface, so can have a
boundary layer and a stellar magnetic field. 

The main premise throughout this paper is that progress in
understanding accretion in {\em any} of these objects should give us
some pointers to understanding accretion in {\em all} of
them. Galactic sources are intrinsically less luminous, but a great
deal closer than the AGN, so generally are much brighter. The
variability timescales also scale with the mass of the central object
so it is much easier to study {\em changes} in the accretion flow onto
Galactic sources. Thus we can use the Galactic sources as a
laboratory for understanding accretion processes, and then see
how much of this can be transfered to AGN. 

\section{Instabilities in Accretion Flows}

Accretion onto a black hole via an optically thick disk has been known
for decades to produce rather robust spectral predictions (Shakura \&
Sunyaev 1973; hereafter SS). The emission should consist of a sum of
quasi--blackbody spectra, peaking at a maximum temperature of $\sim
0.6$ keV for accretion rates around Eddington onto a $\sim 10M_\odot$
GBH, or at $\sim 0.3$ keV for
accretion rates at $\sim 1$ per cent of Eddington. The whole 
accretion disk structure is determined mainly by the luminosity
(i.e. mass accretion rate) as a fraction of the Eddington luminosity,
$\dot{m}$, and is only weakly dependent on the mass of the black hole. 

However, these steady state accretion disk solutions are generically
unstable. At low mass accretion rates then the point at which hydrogen
makes the transition between being mostly neutral to mostly ionised
gives a dramatic instability. When a part of the disk starts to reach
temperatures at which the Wien photons can ionised hydrogen then these
photons are absorbed. This energy no longer escape from the disk, so
it heats up, which produces higher temperatures and more photons
which can ionise hydrogen, so more absorbtion and still higher
temperatures. The runaway heating only stops when hydrogen is mostly
ionised. 

If the temperature crosses the hydrogen ionisation instability
point at {\em any} radius in
the disk then the {\em whole disk} is unstable. This classic disk
instability (DIM) is responsible for a huge variety of variability
behaviour. In the neutron stars
and black holes the DIM is made more complex by X--ray irradiation:
the outburst is triggered by the DIM, then irradiation contributes to
the ionisation of hydrogen, so controls the evolution of the disk,
while it can also enhance the mass transfer from the companion star
(e.g. the review by Lasota 2001).

There may also be a further instability at high mass accretion rates,
where the pressure in the disk is dominated by radiation. If the
heating rate scales with the total pressure, then this scales as
$T^4$ where radiation pressure dominates.
A small increase in temperature leads to a large increase in
heating rate, and so to a bigger increase in temperature. The
runaway heating only stabilises when the timescale for the radiation to
diffuse out of the disk is longer than the accretion timescale for it
to be swallowed by the black hole (optically thick advection:
Abramowicz et al. 1988). The spectrum of these slim
disks are slightly different to those of
standard SS disks, as the energy generated in the innermost orbits is
preferentially advected rather than radiated (Watarai \& Mineshige
2002). At these high mass accretion rates the disk can also overheat, 
causing the inner disk to become strongly comptonised
(Shakura \& Sunyaev 1973; Beloborodov 1998).

However, there is much uncertainty about the disk structure at such
high mass accretion rates. The association of the viscosity with the
magnetic dynamo means that the viscous heating may scale only with the
gas pressure, rather than the total pressure, which removes the
radiation pressure instability (Stella \& Rosner 1984). 
Even the advective cooling may be 
circumvented if the disk becomes clumpy, so that the disk material is
not as efficient in trapping the radiation (Turner et al., 2002;
Gammie 1998; Krolik 1998).

Figure 1a shows the equilibrium solutions (i.e. places
where heating balances cooling) of an SS disk at a given radius
and viscosity. The disk surface density, $\Sigma=\int \rho
dz \propto \tau$, where $\tau$ is the optical depth of the disk, 
generally increases as the mass accretion rate increases, but
the hydrogen ionisation and radiation pressure instabilities are so
strong that the disk surface density actually decreases at these
points. The dotted
line shows where the disk behaviour becomes very uncertain. 

\section{Observations of Accretion Flows in Galactic Black Holes}

Most black hole binaries are in systems where the outer edge of the
disk is cool enough to dip below the hydrogen ionisation point. They
are generically transient, and show large variability. These give us a 
sequence of spectra at differing mass accretion rates onto the central
object, allowing us to test accretion models. 

What is seen is generally very different to the simple Shakura-Sunyaev
disk emission ideas outlined above.  At high mass accretion rates (approaching
Eddington) the spectra are dominated by a soft component at $kT\sim 1$
keV which is strongly (very high state: VHS) or weakly (high state:
HS) Comptonized by low temperature thermal (or quasi--thermal)
electrons with $kT\sim 5-20 $ keV (Zycki et al., 1998; Gierlinski et
al., 1999; Kubota et al., 2001). 
There is also a rather steep power law
tail ($\Gamma\sim 2-3$) which extends out beyond $511$ keV in the few
objects with good high energy data (e.g. Grove et al., 1998). At lower
mass accretion rates, below $\sim 2-3$ per cent of Eddington, 
there is a rather abrupt
transition when the soft component drops in temperature and
luminosity. Instead this (low state: LS) spectrum is dominated by
thermal Comptonization, with $\Gamma < 1.9$, rolling over at energies
of $\sim 150$ keV (see e.g. the reviews by Tanaka \& Lewin 1995; van
der Klis 1995; Nowak 1995).  This spectral form seems to continue even
down to very low luminosities ($\sim 0.01$ per cent of Eddington: the
quiescent or off state e.g. Kong et al. 2000).  Figure 1b shows a
selection of VHS, HS and LS spectra from the GBH transient RXTE
J1550-564.

   \begin{figure}
\begin{tabular}{cc}
{\psfig{file=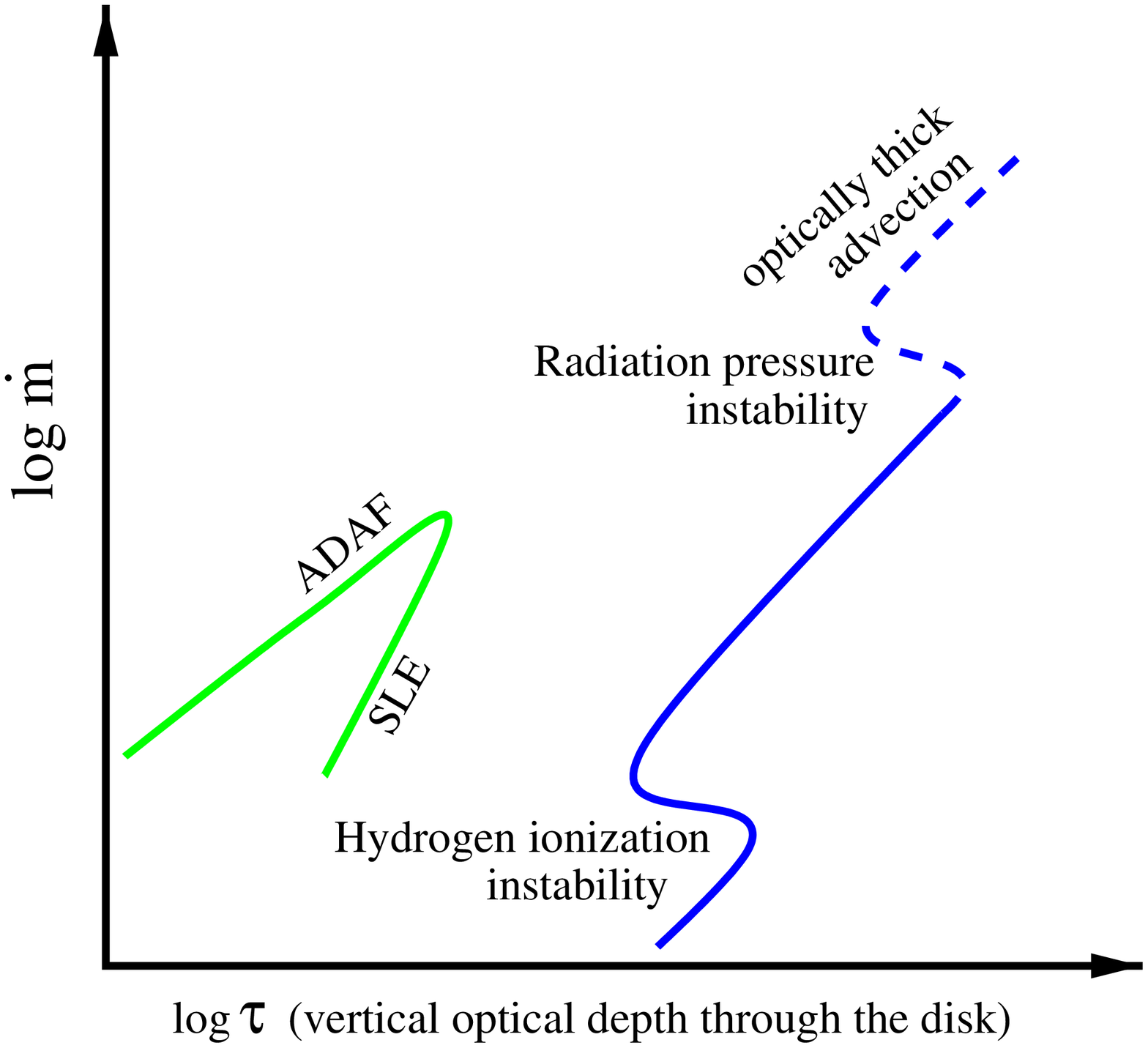,width=0.45\textwidth}}
&
{\psfig{file=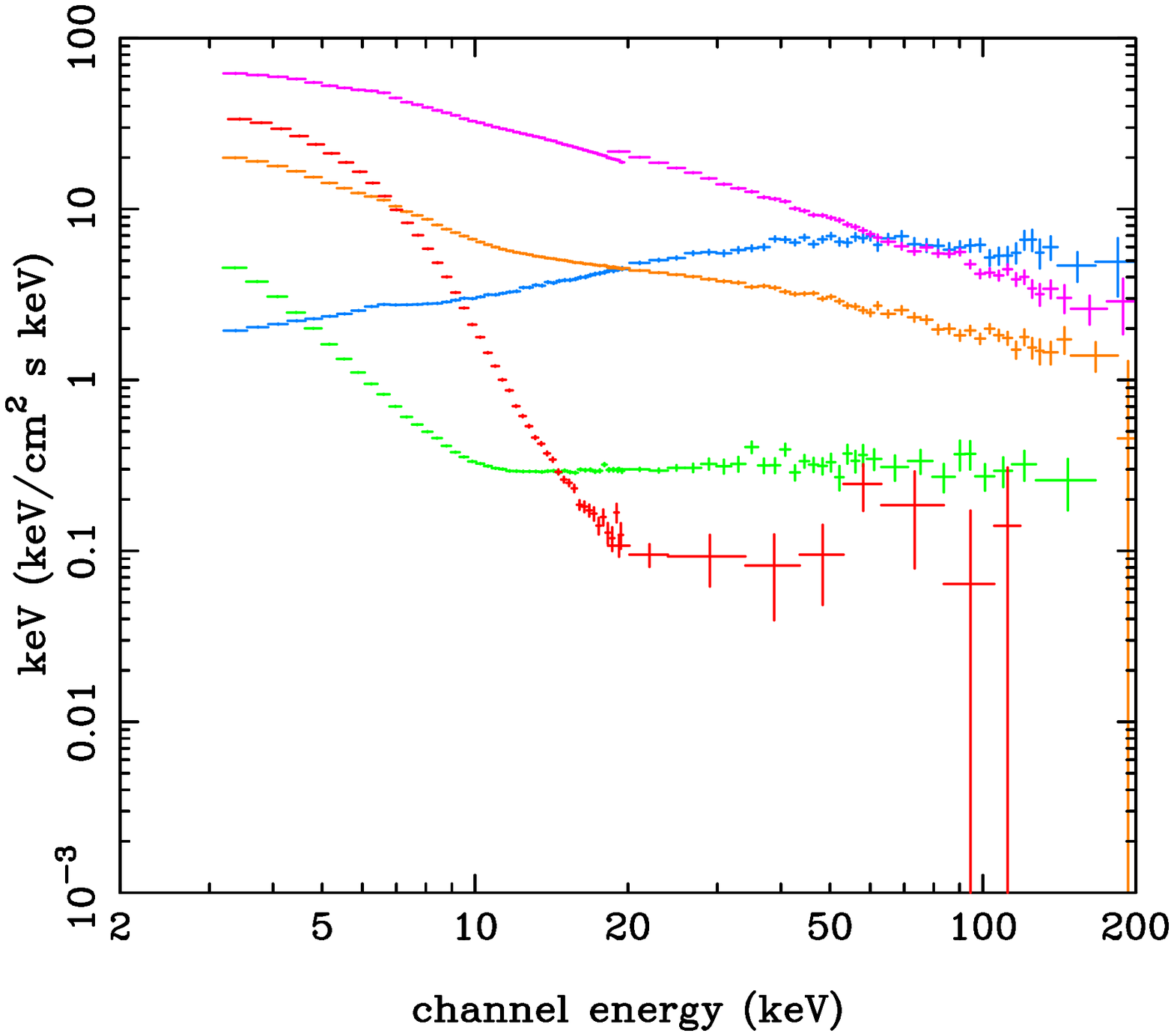,width=0.45\textwidth}}
\end{tabular}
   \caption{
(a) The multiple solutions to the accretion flow
equations. The right hand line is the Shakura-Sunyaev disk, modified
by advective cooling at the highest mass accretion rates, and by
atomic opacities at the lowest mass accretion rates. The left hand
line is the optically thin flows, the advective flows (ADAF) and
Shapiro-Lightman-Eardley (SLE) solutions. 
(b) RXTE PCA and HEXTE data from RXTE J1550--564 showing the
Low state (peaking at 100 keV), and a variety of high mass acccretion
rate spectra, all of which peak at $\sim 1$ keV. There is the high
state (disk sharply peaking at 1 keV), ultrasoft spectrum (disk showing 
rounded peak at 1 keV) and two high states (disk strongly comptonised to
give a smooth, steep spectrum). 
}
\end{figure}

While XMM-Newton and Chandra have opened up new windows in high
resolution X-ray imaging and spectroscopy, RXTE gives an unprecedented
volume of data on these sources.  To get a broad idea of the range of
spectral states then quantity as well as quality is
important!  But this also means that plotting individual spectra is
too time consuming. Colour-colour and colour-intensity diagrams have
long been used in neutron star X-ray binaries to get an overview of
source behavior. The problem is that they often depend on the
instrument response (counts within a certain energy range) and on the
absorbing column towards the source. To get a measure of the
source behaviour we want to plot {\em intrinsic} colour,
i.e. unabsorbed flux ratios over a given energy band. To do this we
need a physical model. Plainly there can be emisison from an accretion
disk, together with a higher energy component from
comptonisation. Reflection of this emission from the
surface of the accretion disk can also contribute to the
spectrum. Thus we use a model consisting of a multicolour accretion
disk, comptonised emission (which is {\em not} a power law at energies
close to either the seed photon temperature or the mean electron
energy), with gaussian line and smeared edge to roughly model the
reflected spectral features, with galactic absorption.

We use this model to fit the RXTE PCA data from 
several different black holes from archival RXTE data
to follow their broad band spectral evolution.
We choose 4 energy bands, 3-4 keV, 4-6.4 keV, 6.4-9.7 keV
and 9.7-16 keV, and integrate the unabsorbed model over these ranges
to form {\em intrinsic} colours and use the generally fairly well
known distance to convert the extrapolated bolometric flux to total
luminosity. Again, since the mass of the central object is fairly well
known we can translate the bolometric luminosity into a fraction of
the Eddington luminosity.

Figure 2a shows a hard colour versus luminosity plot for the black holes
Cyg X-1 (diamonds), GX339-4 (squares) and J1550-564 (circles).  
The general trend is spectra with $L/L_{Edd}<0.03$ to be hard while those
at higher luminosities are soft. The hard spectra from J1550-564 at 10
per cent of Eddington are from the extremely rapid 
{\em rise} to outburst (see Wilson \& Done 2001) where the
accretion flow was presumably far from steady state. Apart from this,
it is clear that there is a transition from hard to soft spectra at
a few per cent of Eddington. 

The huge range in spectral
behaviour is brought out more clearly on a
colour-colour plot (Figure 2b). The hardest spectra (low/hard state)
form a well defined diagonal track, where hard and soft colour change
together while the soft spectra show an amazing variety of shapes.
The points corresponding to the spectra in figure 1b are
(2,1.5) for the low/hard, (0.7,0.7) for the high state, (1.4,0.8) for
the VHS and (1,0.1) for the ultrasoft spectrum.

   \begin{figure}
\begin{tabular}{cc}
{\psfig{file=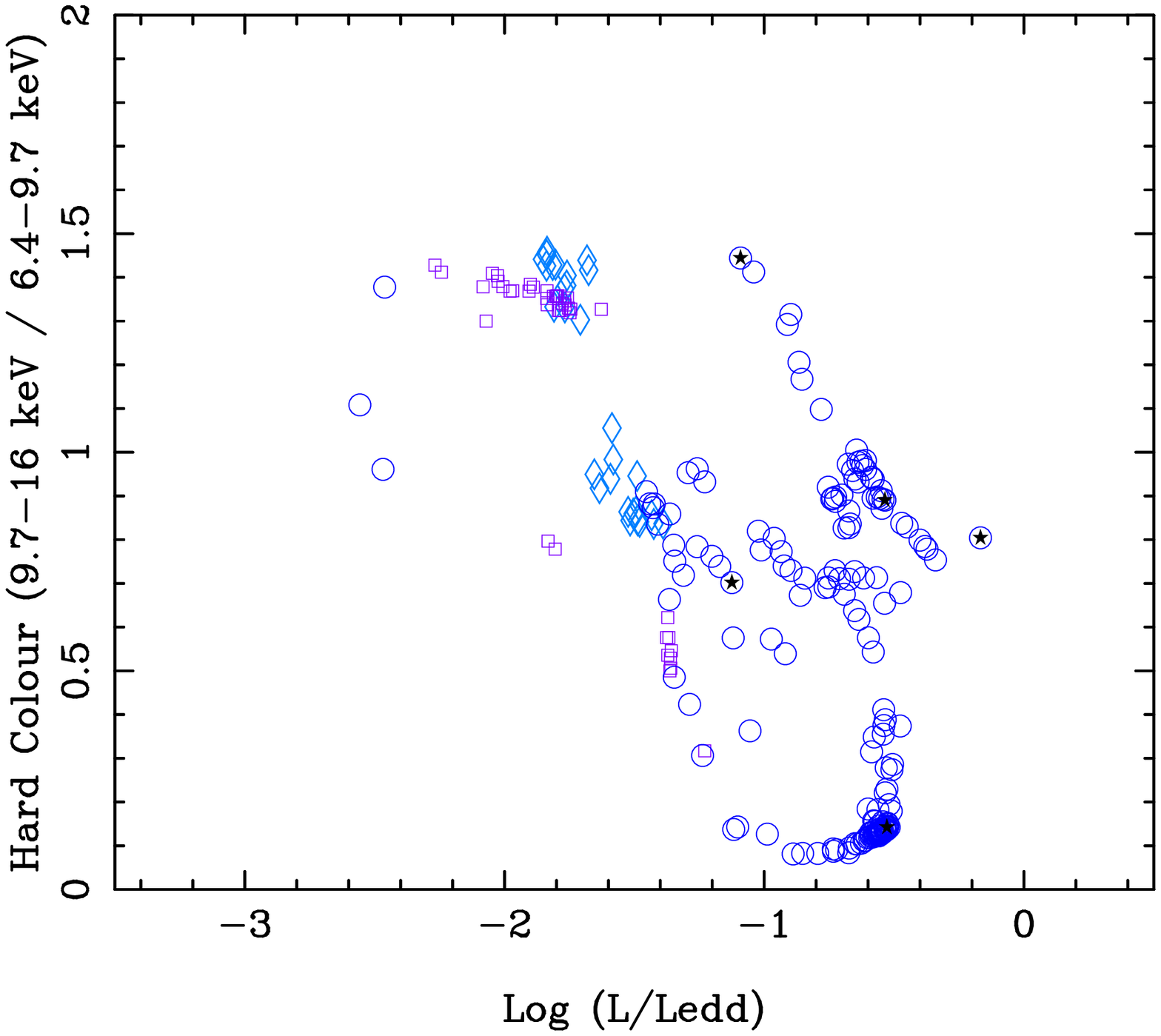,width=0.45\textwidth}}
&
{\psfig{file=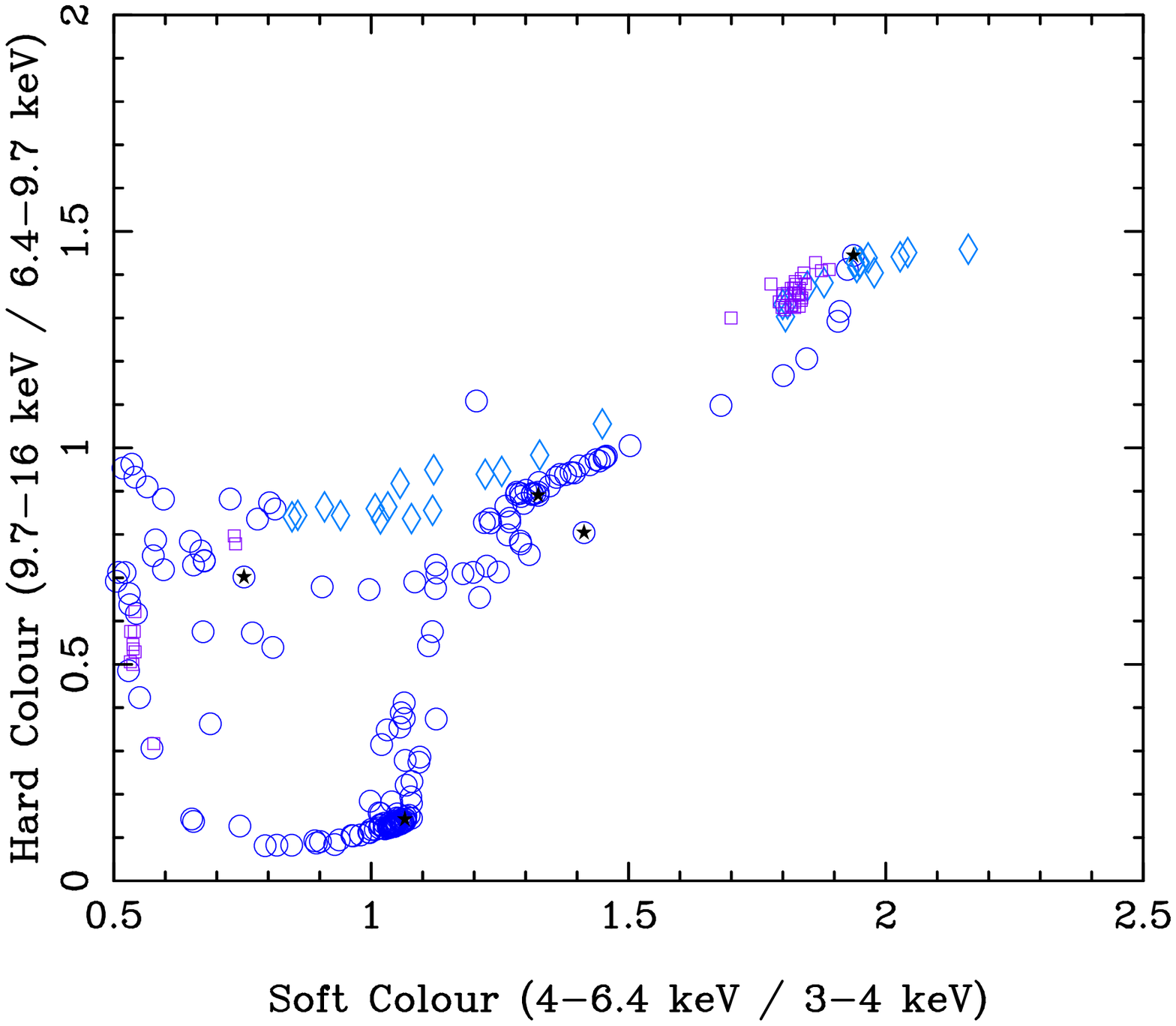,width=0.45\textwidth}}
\end{tabular}
\caption{Luminosity-colour and colour-colour plots for the
RXTE PCA data on the Galactic black holes GX339 (squares), Cyg X-1
(diamonds) and J1550-564 (circles). Points corresponding to the
spectra in Figure 1b are marked with stars.}
\end{figure}

\section{Observations of Accretion Flows around Neutron stars}

Neutron stars without a strong magnetic field ($B<10^{12}$ G)
come in two flavors, named atolls and Z sources. Z sources are named
after a Z-shaped track they produce on an X-ray colour-colour diagram,
while atolls are named after their C (or atoll) shaped track. 
These differences between the 
two LMXBs categories probably reflect differences in both mass 
accretion rate, $\dot{M}$, and magnetic field, $B$, with the Z 
sources having high luminosity (typically more than 50 per cent of 
the Eddington limit) and magnetic field ($B \ge 10^9$ G) while the 
atolls have lower luminosity (generally less than 10 per cent of 
Eddington) and low magnetic field ($B \ll 10^9$ G) (Hasinger \& van 
der Klis 1989). 
 
Most of these are {\em stable} to the
disk instability (neutron stars round the same mass
companion need to be closer than a black hole for the star to fill its 
Roche lobe. The disk is smaller, and hotter so less likely to trigger
the hydrogen ionisation instability: King \& Ritter 1998). However,
there are a few (probably evolved) atoll systems which are transient,
Figure 3a shows the colour-luminosity 
plot for these systems (small filled circles), 
while figure 3b shows their colour-colour
diagram. Plainly there is again a switch from a well defined hard
state to a softer spectrum at $\sim 10$ \% of Eddington, 
although here the soft state also forms a single well defined track.
Because these are {\em intrinsic} colours, this diagram can be
overlaid on that for the black holes. It is clear that the neutron
star spectra evolve in very different ways with mass accretion rate, 
and that the softest black hole spectra (high state and ultrasoft) are
not seen in these atoll sources. 

\begin{figure}
\begin{tabular}{cc}
{\psfig{file=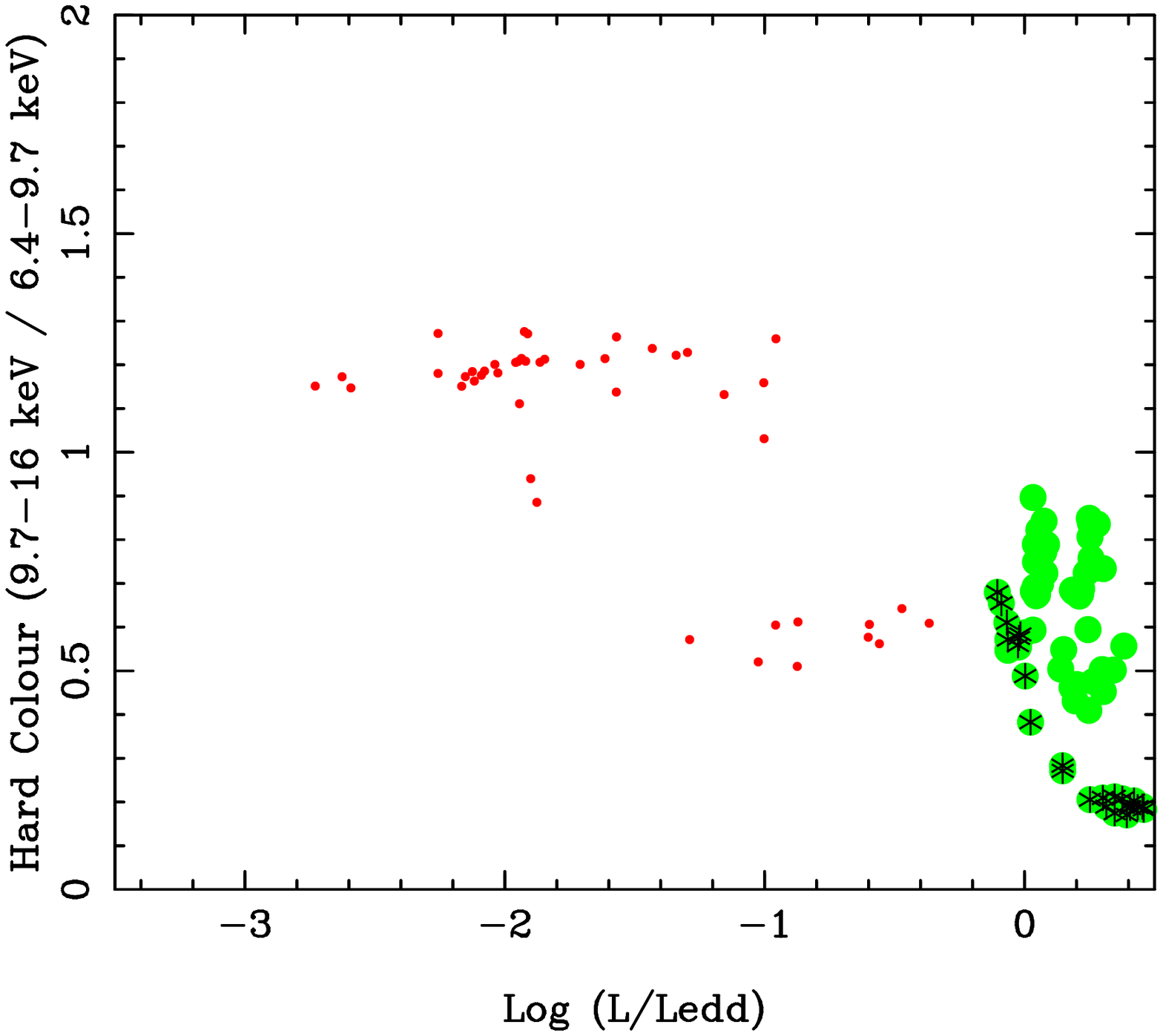,width=0.45\textwidth}}
&
{\psfig{file=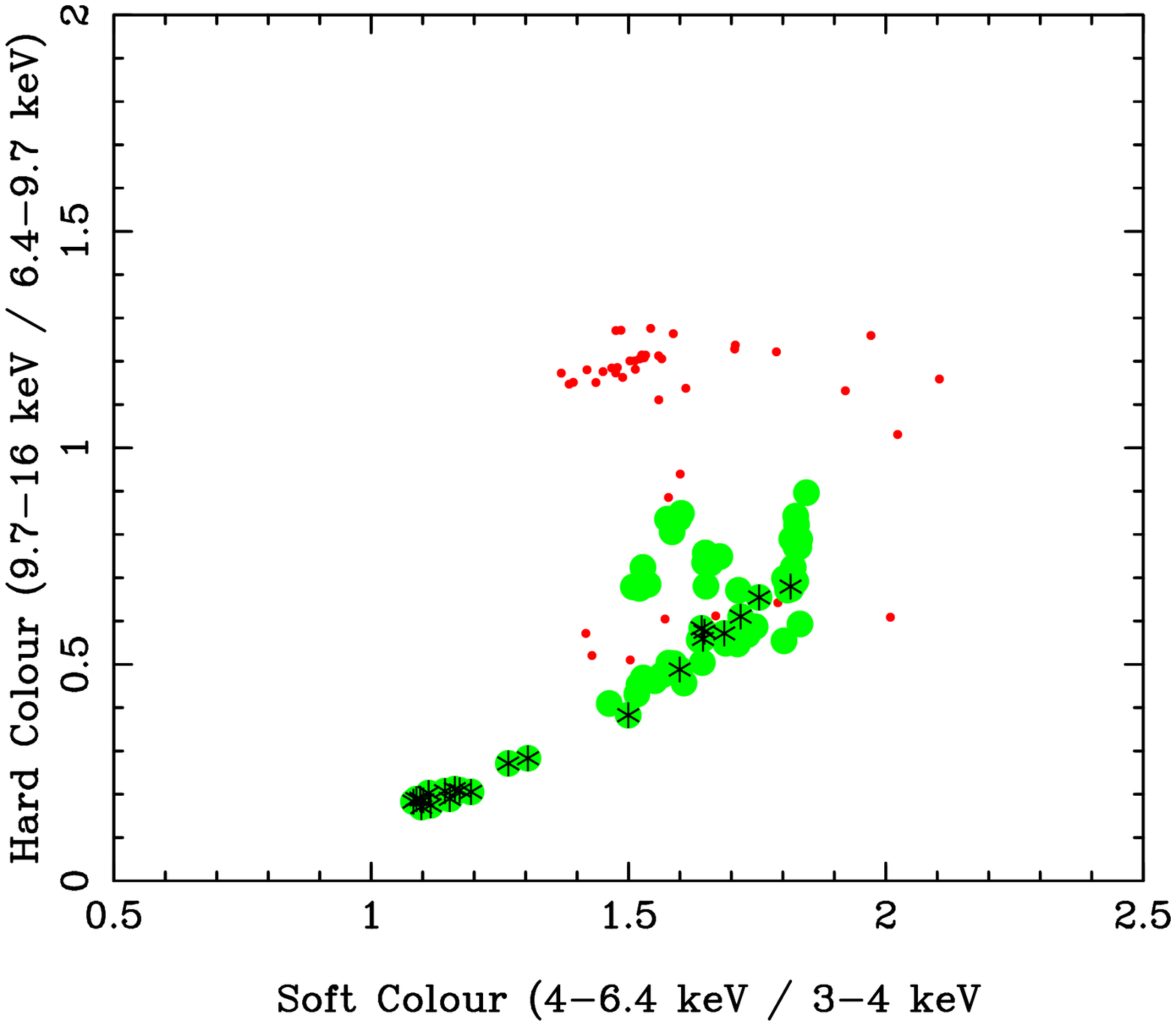,width=0.45\textwidth}}
\end{tabular}
\caption{Luminosity-colour and colour-colour plots for the
RXTE PCA data on the transient atolls (low magnetic field disk
accreting neutron stars - small dots). 
Again there is a hard to soft spectral switch
at $\sim 10$ per cent of Eddington. The standard Z sources Cyg X-2,
Sco X-1 and GX 17+2 are shown as large circles, while stars show Cir X-1.
}
   \end{figure}

However, the 'ultrasoft' spectrum is not a unique black hole signature
as the odd neutron star binary Cir X--1 (tentatively classed as a Z
source) has been known for a long time to show such a
spectrum. Figures 3a and b also include data from Cir X-1 (stars) and
standard Z sources (large filled circles).  It is immediately
clear that the standard Z sources do not vary by much from Eddington
and that their spectral variability along their Z is rather small also
(the Z track in figure 3b is mostly masked by the symbol size!). And
while Cir X-1 is not convincingly like either class of system, the
important point here is that Cir X-1 certainly ends up at the same
'ultrasoft' spectrum as the black holes on a colour-colour diagram.
Perhaps when the mass accretion rate is extremely high then the flow
is so optically and geometrically thick that it completely swamps the
central object, so that the nature of the central object becomes
unimportant. 
However, {\em no} neutron star systems show the high
state spectra. These may be a unique black hole signature, as the
boundary layer which should always be present in the neutron stars
will provide a higher temperature (harder) component, unless the flow
is so thick that it completely buries the central object.  

\section{X-ray emission in the low/hard state}

Hard X--rays are a generic feature of accretion onto a black hole in
all spectral states, yet the standard SS disk models cannot produce
such emission. Even worse, the observations show that we have to
explain different sorts of hard X--ray emission, with a
fairly well defined spectral switch between the hard and 
soft spectra at luminosities of around 3 per cent of Eddington. 
This mechanism has to work also in the low magnetic field neutron
stars at low mass accretion rates. 

To get hard X-rays then a large fraction of the gravitational energy
released by accretion must be dissipated in an optically thin
environment where it does not thermalise, and so is able to reach much
higher temperatures. For the SS disk, an obvious candidate is that
there are magnetic flares above the disk, generated by the
Balbus--Hawley MHD dynamo responsible for the disk viscosity 
(Balbus \& Hawley 1991). 
Buoyancy could cause the magnetic
field loops to rise up to the surface of the disk, so they can
reconnect in regions of fairly low particle density. This must be
happening at some level, and is shown by numerical simulations,
but current models (although these are highly incomplete as in
general the simulated disks are not radiative)
do not carry enough power out from the disk to reproduce the observed
low/hard state (Miller \& Stone 2000).

If such mechanisms produce the low state spectra, then we also need a
mechanism for the transition. The obvious candidate is the radiation
pressure instability. However, the disk is no longer a standard SS
disk as much of the power is dissipated in a magnetic corona rather
than in the optically thick disk.  The disk is cooler and denser
(Svensson \& Zdziarski 1994), which means that it is gas pressure
dominated up to higher mass accretion rates than a standard SS disk.
The disk just starts to hit the radiation pressure instability at a
few per cent of Eddington when the fraction of power dissipated in the
corona is around 60 per cent, but in the limit where all the power is
dissipated in the corona then the disk is {\em stable} at all mass
accretion rates below the Eddington limit (Svensson \& Zdziarski 1994).

An alternative to the SS disk models is that the inner disk is
replaced by an optically thin, X--ray hot accretion flow.  This flow
still has to dissipate angular momentum, so magnetic reconnection is
still the source of heating, but the assumption here is that the
accretion energy is given mainly to the protons, and that the
electrons are only heated via Coulomb collisions. The proton
temperature approaches the virial temperature so pressure support
becomes important and the flow is no longer geometrically thin. The
electrons cool by radiating, while the protons cool only by Coulomb
collisions, so the flow is intrinsically a two temperature
plasma. Where the electrons radiate most of the gravitational energy
through comptonisation of photons from the outer disk then the
solution is that of a Shapiro-Lightman-Eardley (SLE: Shaprio et al
1976) flow, while if the protons carry most of the accretion energy
into the black hole then this forms the advection dominated accretion
flows (Narayan \& Yi 1996). These two solutions are related, as in
general both advection and radiative cooling are important (Chen et al
1995, Zdziarski et al 1998). They are sketched as the grey line in
Figure 1a.  The SLE flows are unstable, although not dramatically so
(Zdziarski 1998), while ADAFs are stable (Narayan \& Yi 1995).

Importantly, both optically thin flows give
typical electron temperatures of $\sim 100$ keV, as required to
explain the low/hard spectra, and both can only exist (in fact they
merge) at mass
accretion rates below a critical value of a few per cent of Eddington
(Chen et al., 1995) as when the flow becomes
optically thick then the Coulomb collisions become efficient so that
the flow collapses back into the one temperature SS disk solution.
Hence these flows could produce both the quiescent
and LS spectra, and the collapse of such flows may give the physical
mechanism for the hard-soft state transition (Esin et al., 1997)

\section{X-ray emission in the high/soft states}

The high mass accretion rate spectra are dominated by the disk
emission, but there is always a weak, power law X-ray tail (HS) and
sometimes strong comtonisation of the disk emission (VHS). 
Magnetic reconnection above the disk is
really the only known contender for producing the X-ray tail,
but here the electrons must have a nonthermal spectrum
(Gierlinski et al., 1999). The strong comptonisation of the disk may
be connected to this same mechanism, with the electrons partailly
thermalising to produce a thermal/nonthermal hybrid plasma
(Poutanen \& Coppi 1998). Alternatively, the comptonisation may be
connected to the inner portion of the disk, either overheating
in the standard disk equations (SS; Beloborodov 1998), or as a result
of trying to change its structure when it reaches the radiation
pressure instability (Kubota et al., 2001).

Figure 4 sketches all these potential hard X--ray emission mechanisms.

\begin{figure}
\centerline{\psfig{file=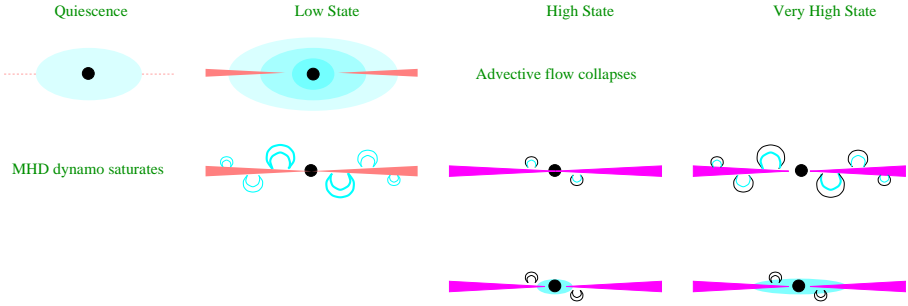,width=0.95\hsize}}
\caption{Potential X--ray emission mechanisms in the various spectral
states. In quiescence the disk is NOT in steady state (as indicated by
the dotted line). Hydrogen is mostly neutral and the MHD dynamo
probably cannot operate. In the low state the accretion flow is in 
(quasi)steady state. Hydrogen is ionised so the MHD dynamo works, and
if there is reconnection above the disk then this could heat the electrons.
Alternatively, the X--ray emission in 
both quiescence and and low state could be powered by an advective
flow. In the high and very high states the only serious contender for
the hard power law tail is magnetic reconnection leading to a 
non--thermal electron distribution (indicated by the black loops),
while the thermal electrons could be part of the same mechanism (grey
loops) or could be associated with the inner disk.}
\end{figure}

\section{A Unified description of Accretion flows in X-ray binaries}

A truncated disk/inner hot flow model can explain the evolution of the
different classes of X-ray binaries on a colour-colour and
colour-luminosity diagram. For all types of sources we assume that the
evolution of the source can be explained if the main parameter driving
the spectral evolution is the average mass accretion rate, $\dot{m}$. 
This is {\em not} the same as the instantaneous mass accretion rate,
inferred from the X-ray luminosity. There is some much longer
timescale in the system, presumably tied to the response of the disk
and/or inner flow (e.g. van der Klis 2000; 2001). While the truncation
mechanism is not well understood, it seems likely that conductive
heating of a cool disk by the hot plasma (especially if they are
magnetically connected) could lead to evaporation of the disk
(Meyer \& Meyer-Hofmeister 1994; Rozanska \& Czerny 2000), so that
there is a smooth tranistion from an outer disk to inner hot flow.

A qualitative picture could be as follows, starting at low $\dot{m}$.
For the black holes, as $\dot{m}$ increases then the truncation radius
of the disk decreases, so it penetrates further into the hot flow. The
changing geometry gives more soft photons to cool the hot flow, and so
leads to softer spectra. Both hard and soft colours soften together
as the power law produced by compton scattering in the inner flow
is still the only component within the PCA bandpass. 
Then the inner flow reaches its maximum luminosity, and collapses. As
the inner disk replaces the hot flow then the 
disk is close enough to contribute to the spectrum above 3
keV, so the soft colour abruptly softens. The hard X-ray 
tail is produced by the small fraction of
magnetic reconnection which takes place outside of the optically thick
disk (high state). At even higher $\dot{m}$ then the disk structure is
not well understood, but there seems to be a
choice of two disk states, one in which the inner disk 
emission is strongly comptonised (very high state), characterised by
moderate hard and soft colours or else is extremely disk dominated, as
characterised by very soft hard colours. 

The atoll neutron stars show similar evolution, except that they also
have a solid surface and so have a boundary layer.  At low $\dot{m}$
then the disk is truncated a long way from the neutron star, and the
boundary layer is mostly optically thin, so it joins smoothly onto the
emission from the inner accretion flow.  Reprocessed photons from the
X-ray illuminated surface form the seed photons for compton cooling of
the inner flow. As $\dot{m}$ increases, the disk starts to move
inward, but the cooling is dominated by seed photons from the neutron
star rather than from the disk, so the geometry and hence the high
energy spectral shape does not change. The atolls keep constant hard
colour, but the soft colour softens as the seed photon energy moves
into the PCA band. The inner flow/boundary layer reaches its maximum
luminosity when it becomes optically thick. This causes the hard
colour to soften as the cooling is much more effective as the boudary
layer thermalises, so its temperature drops. The disk replaces the
inner hot flow, so the disk temperature starts to contribute to the
PCA bandpass so the soft colour softens abruptly.  The track moves
abruptly down and to the left during this transition.  After this then
increasing $\dot{m}$ increases the disk temperature, so the motion is
to higher soft colour (Gierlinski \& Done 2002a; 2002b).  The 'banana
state' is then analogous to the high/soft state in the galactic black
holes, but with additional luminosity from the boundary layer.

The Z sources can be similar to the atolls, but with the addition of a
magnetic field (Hasinger \& van der Klis 1989). 
In general they are stable to the disk instability
because of the high mass accretion rates, so vary only within a factor
of 2. The disk evaporation  efficiency 
decreases as a function of increasing mass accretion rate, so this 
cannot truncate the disc in the Z sources. Instead the truncation is 
likely to be caused by stronger magnetic field, but here the 
increased mass accretion rate means that the inner flow/boundary layer
is  already optically thick, and so cooler (Gierlinski \& Done 2002a).

Models with a moving inner disk radius at low mass accretion rates
also can qualitatively explain the variability power spectra of these
sources. They show characteristic frequencies in the form both of
breaks and Quasi Periodic Oscillations (QPO's). These features are
related ($f_{break}\sim 10 f_{QPO}$ for the low frequency QPO), and
they {\em move}, with the frequencies generally being higher
(indicating smaller size scales) at higher $\dot{m}$ (see e.g. the
review by van der Klis 2000).  Recent progress has concentrated on the
similarity between the relationship between the QPO and break
frequencies in black holes {\em and} neutron star systems (e.g. the
review by van der Klis 2000)
If they truly are the same phenomena
then the mechanism {\em must} be connected to the accretion disk
properties and not to the magnetosphere or surface of the neutron
star. While the variability is not yet understood in detail, {\em all}
QPO and break frequency models use a sharp transition in the accretion
disk in some form to pick out a preferred timescale (e.g. van der Klis
2000), so by far the easiest way to change these frequencies is to
change the inner disk radius.

\section{Observational tests of the Low/hard state geometry}

While disk truncation with an inner X-ray hot flow
can form a coherent picture for the low mass
accretion rate X-ray binaries and Z sources, there is the alternative 
model in which the low state is produced by magnetic flares above an
untruncated disk. This is a very different geometry to that of a
truncated disk, so there should be some testable, 
observational signatures of the inner disk which can enable us to
discriminate between these two models.

\subsection{Direct disk emission}

The most direct way to see whether there is an inner disk is look at
the disk emission. The SS disk models make clear predictions about the
temperature and luminosity of the disk, and these can easily be
modified for the case where some fraction $f$ of the energy is dissipated
above and below the disk (Svensson \& Zdziarski 1994). Both disk 
temperature, and luminosity, and fraction of energy emitted in the corona
can be {\em observed} with a broad band spectrum showing both
disk and hard X-ray emission, so the disk inner radius can be constrained
directly from the data. 

However, the expected temperatures for disks around black holes
accreting at a few per cent of Eddington are in the EUV/soft X-ray
region where interstellar absorption is important. Most black holes
and neutron stars are in the galactic plane, so there is high
obscuration.  ASCA and SAX observations of Cyg X-1 in the low/hard
state infer disk temperatures of $\sim 0.1$ keV (di Salvo et
al., 2001), while the broad band SAX spectrum
also shows its total bolometric luminosity was about 2 per cent of
Eddington, with 70 per cent of this
emitted in the hard X-ray spectrum (di Salvo et al., 2001).
Both energetics and disk temperature are consistent with a disk
truncated at about 50 Rs, but an untruncated disk with 
70 per cent of the power dissipated above the optically thick disk
material has a temperature of $\sim 0.2$ keV. This is marginally
inconsistent with that observed, but not dramatically so, and the
absorption to Cyg X-1 is fairly high so there are some systematic
uncertainties. 

The disk is much more clearly seen in the transient black hole system
RXTE J1118+480. This has extremely low galactic absorption, so the
disk emission can be detected with HST and EVUE.  The temperature is
$\sim 0.02$ keV, while the luminosity is again a few per cent of
Eddington, with about 40 per cent of the total emitted in hard
X-rays. This low disk temperature is completely inconsistent with a
disk extending down to the last stable orbit (McClintock et
al. 2001), even with 90 per cent of the power dissipated in a hot
corona. Esin et al., (2001) sucessfully fit the overall spectral shape
with a truncated disk/inner advective flow model (Figure 5).

\begin{figure}
\centerline{\psfig{file=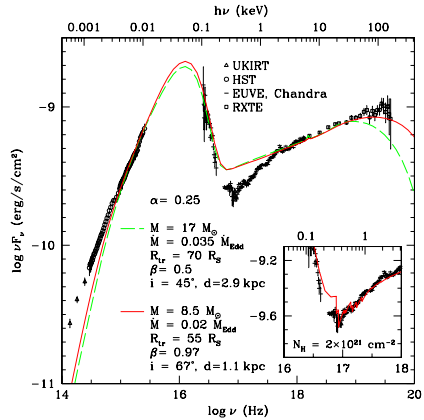,width=0.45\hsize}}
\caption{The multiwavelength broad band spectrum of the black hole
transient RXTE J1118+480, from Esin et al., (2001). The low galactic
column to this source means the direct disk emission can easily be
observed with HST and EUVE.}
\end{figure}

\subsection{Reflection}

An independent way to study the extent of the optically thick
accretion flow is via reflection.
Wherever hard X--rays illuminate optically thick
material then there is some probability that the X--rays can be
reflected. The reflection probability is given by a
trade--off between the importance of electron scattering and
photo--electric absorption.  Since the latter is energy dependent, the
albedo is also energy dependent, with higher energy photons being
preferentially reflected due to the smaller photo--electric opacity of
the material. This gives rise to a reflected spectrum that is harder
than the intrinsic spectrum, with photo--electric edge features,
and associated fluorescent lines imprinted on it. 
Since iron is the highest atomic number element
which is astrophysically abundant, the iron K edge is particularly
prominent, at 7.1--9.3 keV depending on the ionization state of the
reflecting material, together with its associated iron K$\alpha$
fluorescence line emission at 6.4--6.9 keV (see e.g. the review by
Fabian et al., 2000).

The reflected spectrum (line and continuum) is smeared by special and
general relativistic effects of the motion of the disk in the deep
gravitational potential well (Fabian et al., 1989; Fabian et al
2001). The fraction of the incident flux which is reflected gives a
measure of the solid angle of the optically thick disk as observed
from the hard X--ray source, while the amount of smearing shows how
far the material extends into the gravitational potential of the black
hole. This should give a clear test of whether the inner disk is
present. 

The Galactic Black holes in the low/hard state show overwhelmingly
that the solid angle is significantly less than $2\pi$, and that the
smearing is less than expected for a disk extending down to the last
stable orbit (Zycki et al.,1997; 1998; 1999; Gierlinski et al., 1997;
Done \& Zycki 1999; Zdziarski et al., 1999; Gilfanov et al., 1999;
Revnivtsev et al., 2000).  While this is clearly consistent with the
idea that the disk is trunctated in the low/hard state, an alternative
explanation for the lack of reflection and smearing is that the inner
disk or top layer of the inner disk is completely ionised. There are
then no atomic features, and the disk reflection is unobservable in
the 2--20 keV range as it appears instead to be part of the power law
continuum (Ross \& Fabian 1993; Ross et al., 1999, Nayakshin et al
2000).  Ionisation could
be especially important in GBH due to the high disk temperature
predicted by SS untruncated disk models, and indeed in the HS and VHS
the reflection signature from the disk in GBH is clearly ionised
(Gierlinski et al., 1999; Wilson \& Done 2001; Miller et al., 2001).
However, for the low/hard state, the {\em observed} disk temperatures
of less than 0.1 keV are not sufficent to strongly ionise
iron. Collisional ionisation does not strongly distort the
GBH reflection, but photo-ionisation can be very important, and models
with complex photo-ionisation structure can fit the observed
reflection signature by an untruncated disk (Young et al 2001; Done \&
Nayakshin 2001; Ballantyne et al., 2001). However, the inner disk in
such models is not completely invisible as the illuminating photons at
$\sim 100$ keV cannot be reflected elastically. They heat the disk due
to compton downscattering, so it reprocesses and thermalises the hard
X--rays down to temperatures which are typically of order of the
temperature expected from a Shakura-Sunyaev inner disk or
higher. While detailed models of the thermalised emission from complex
ionisation models have not yet been fit to the data, at least 20 per
cent of the illuminating flux should be thermalised by the inner disk,
leading to a higher temperature than observed in RXTE J1118+480.

Reflection and reprocessed, thermal emission from the disk can be
supressed entirely if the magnetic flares are expanding
relativistically away from the disk (Beloborodov 1999). However, the
disk emission in RXTE J1118+480 and Cyg X-1 would then have to be the
intrinsic disk emission, and the expected temperature the observed
luminosity with an untruncated disk is again higher than those
observed.

\section{Conclusions}

A truncated disk at low mass accretion rates is compatible with {\em
all} the constraints on the extent of the accretion disk as measured
by direct emission and reflection. It can also give a unified picture
of the evolution of the broad band spectral shape in both accreting
black holes and neutron stars, and qualitatively explain the
variability power spectra if the disk extends further into the
gravitational potential as the mass accretion rate increases.  
The most convincing single measurement
which challenges alternative, untruncated disk models for the low/hard
state is the observed low temperature disk in the black hole transient
RXTE J1118-480, with supporting evidence for a low disk temperature in
Cyg X-1. Evaporation of the disk into a hot inner accretion flow gives
a plausible physical mechanism for the truncation, and the mass
accretion rate at which the inner flow becomes optically thick gives a
mechanism for the spectral switch from hard to soft spectra. At higher
mass accretion rates the structure of the accretion disk is not well
understood, and the data seem to show a variety of soft spectral
shapes which may be associated with optically thick advection (slim
disks) and/or the radiation pressure instability and/or overheating of
the inner disk. 

We can scale these ideas up to AGN if the accretion flow is simply a
function of radius in terms of Schwarzchild radii and mass accretion
rate in terms of fraction of the Eddington rate.  The disk evaporation
mechanism should still work in the same way around supermassive black
holes (Rozanska \& Czerny 2000).  Thus it seems likely that those AGN
which accrete at a low fraction of the Eddington limit should also be
similar to the low/hard state from galactic black holes. Conversely,
AGN at high mass accretion rates (Narrow line Seyfert 1's and
MCG-6-30-15 ?) are probably the counterparts of the high and very high
state spectra seen in the Galactic black holes where the disk probably
extends down to the last stable orbit.  To test these ideas we again
need some way to determine the extent of the inner disk, but as the
disk temperature scales as $M^{-1/4}$ then it is even harder to
observe the direct disk emission in AGN than in GBH. Reflection is
then probably the best current way to track the extent of the
optically thick accretion disk, and the very broad line seen in
MCG-6-30-15 (Tanaka et al., 1995; Wilms et al., 2002) points to a disk
which extends down to at least the last stable orbit.  However,
results from other AGN are currently controversial.  Lubinski \&
Zdziarski (2001) find that AGN with harder spectra and probably lower
mass accretion rates show less relativistic smearing and less
reflection, consistent with the accretion picture outlined above
for the galactic sources, while previous literature has stressed the
similarity of the line profiles to that of MCG-6-30-15 (e.g. Nandra et
al 1997). Current challenges to observers are to disentangle the
shape of the line in AGN, especially those with hard spectra, so 
as to determine the nature of the accretion flow
in low mass accretion rate AGN, and to theoreticians to 
develop a better understanding of the high mass accretion rate disks. 

\section{Acknowledgements}
Much of this review includes ideas developed in collaborations with
Marek Gierlinski, Aya Kubota and Piotr Zycki.

\smallskip

\end{document}